\documentclass[fleqn,12pt,twoside]{article}
\usepackage{espcrc1}

\def\bq{\begin{eqnarray}}
\def\eq{\end{eqnarray}}
\def\be{\begin{eqnarray}}
\def\ee{\end{eqnarray}}

\usepackage{epsfig}

\newcommand{\AmS}{{\protect\the\textfont2
  A\kern-.1667em\lower.5ex\hbox{M}\kern-.125emS}}

\title{
Generalized parton distributions of few body systems
}

\author{S. Scopetta\address[MCSD]{
Dipartimento di Fisica, Universit\`a degli Studi
di Perugia, via A. Pascoli, 06100 Perugia, Italy
and INFN, sezione di Perugia}}
       
\begin{document}

\maketitle

\begin{abstract}
The relevance of measuring Generalized Parton Distributions
(GPDs) for few nucleon systems is illustrated.
An approach which permits 
to calculate the GPDs of hadrons made of composite constituents
by proper convolutions is described.
The application of the method to the nucleon target,
assumed to be made of composite constituents 
is reviewed. Calculations of GPDs for few
nucleon systems are summarized, with special 
emphasis to the $^3$He target.
\end{abstract}

\vskip 0.5cm


In the last few years \cite{first},
Generalized Parton Distributions (GPDs)
have become one of the most relevant issues
in Hadronic Physics. 
GPDs
enter the non trivial part of
exclusive lepton Deep Inelastic Scattering
(DIS) off hadrons and can be measured, e.g., in
Deeply Virtual Compton Scattering (DVCS),
i.e. the process
$
e H \longrightarrow e' H' \gamma
$ when
$Q^2 \gg m_H^2$
(here and in the following,
$Q^2$ is the momentum transfer between the leptons $e$ and $e'$,
and $\Delta^2$ the one between the hadrons $H$ and $H'$) so that
experimental DVCS programs are taking place.
As observed in \cite{cano1}, 
the knowledge of GPDs of nuclei would permit the study
of their short light-like distance structure, 
and thus the interplay
of nucleon and parton degrees of freedom.
In DIS off a nucleus
with four-momentum $P_A$ and $A$ nucleons of mass $M$,
this information can be accessed in the 
region where
$A x_{Bj} \simeq { Q^2 / ( 2 M \nu) }>1$,
being $x_{Bj}= Q^2 / ( 2 P_A \cdot q )$ and $\nu$
the energy transfer in the laboratory system,
but measurements are difficult, because of 
vanishing cross-sections.
As explained in Ref. \cite{cano1}, 
the same physics can be accessed
in DVCS at lower values of $x_{Bj}$
\cite{cano2}. 
\begin{figure}[ht]
\centerline{\epsfxsize=3.0in\epsfbox{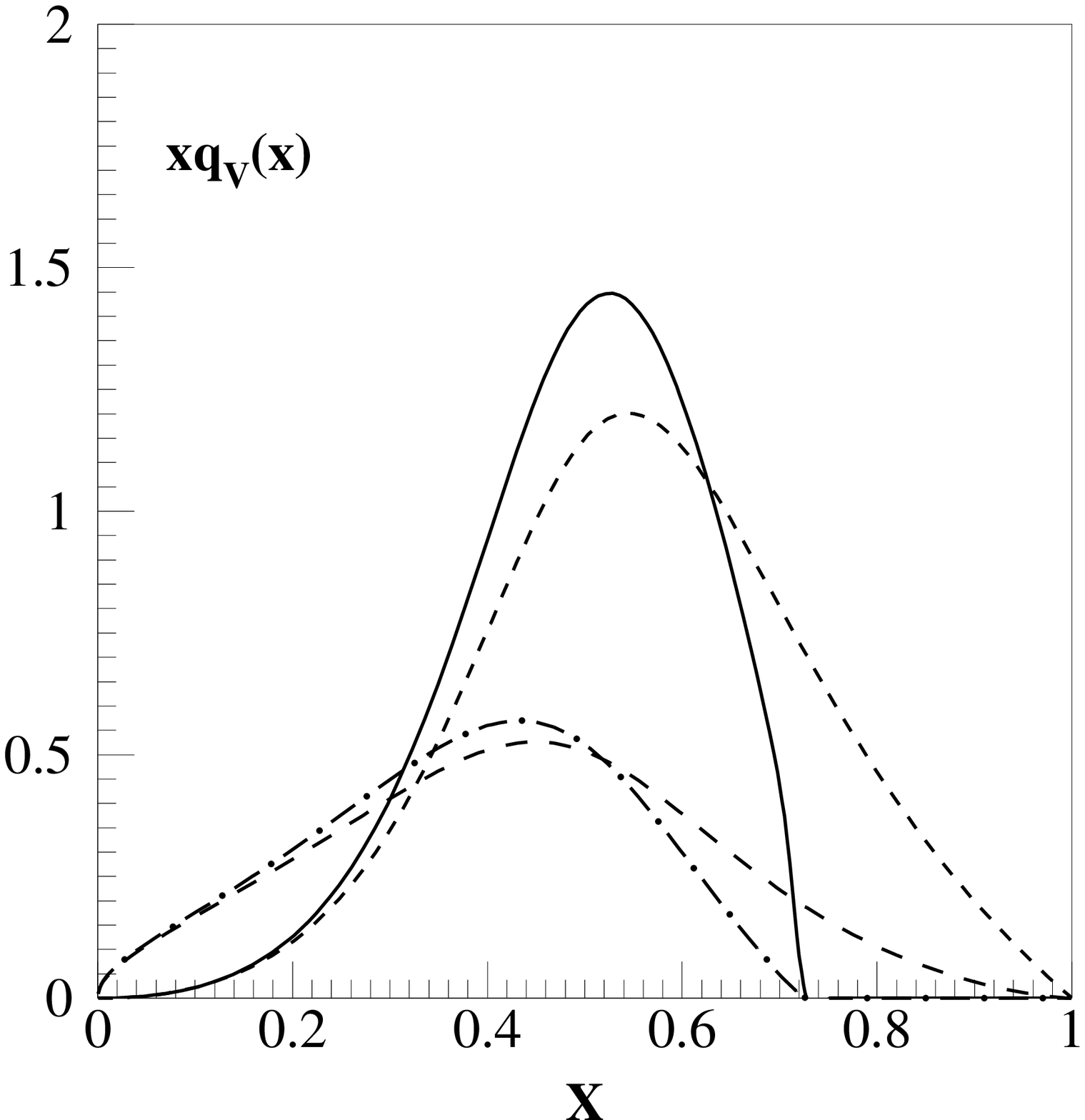}
\epsfxsize=3.0in\epsfbox{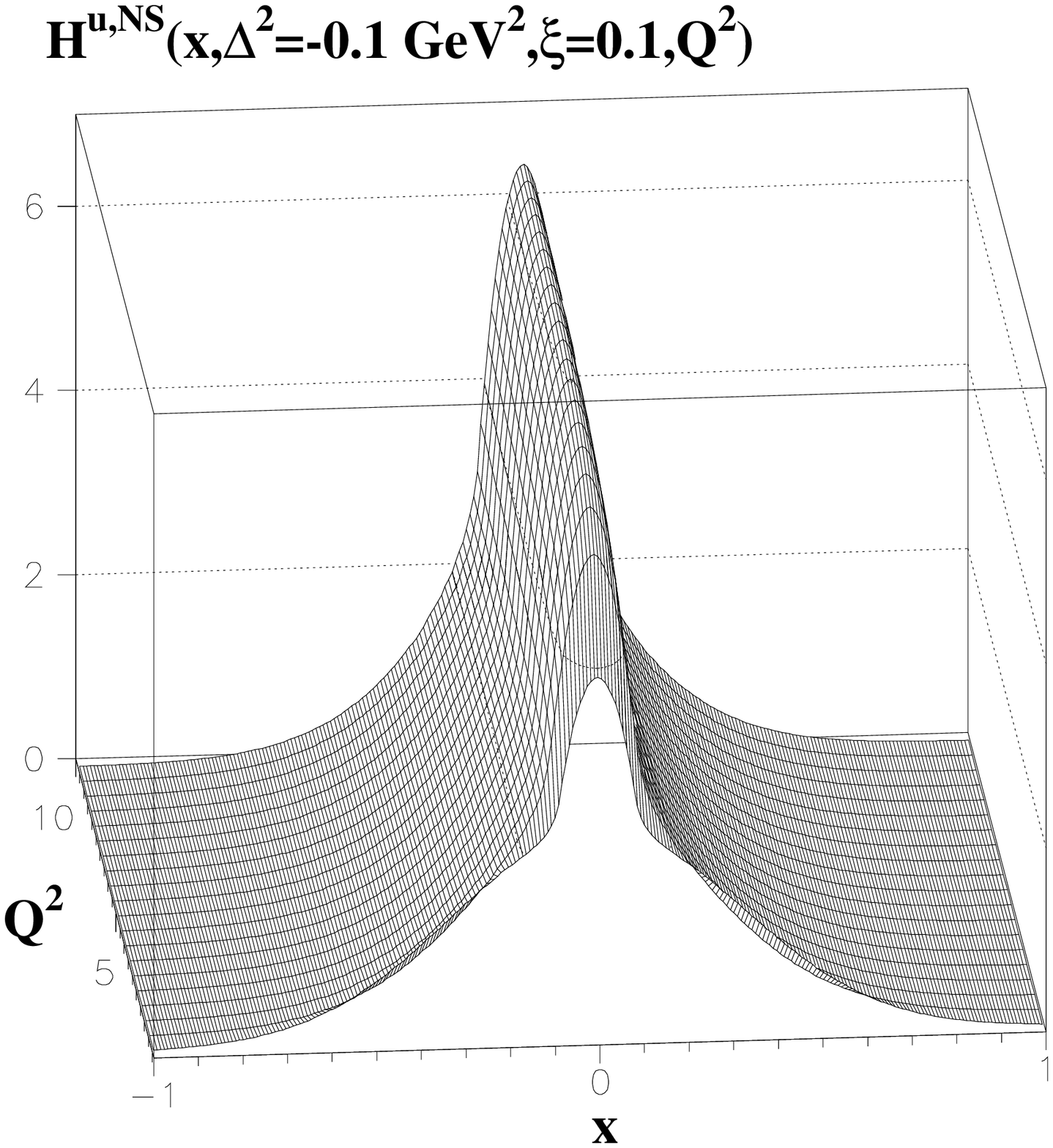}}   
\caption{
Left panel: the valence quark parton distribution in a relativistic
model for a quark-antiquark system (dashed) and its non relativistic
(NR) limit (full); once a structure is introduced
for the constituents, the relativistic and
NR models give the long dashed and dot-dashed results, respectively.
Right panel: The  
GPD $H(x,\xi,\Delta^2)$ for the flavor $u$,
evolved from $\mu_0^2=0.34$ GeV$^2$
to $Q^2$=10 GeV$^2$, 
for $\Delta^2=-0.1$ GeV$^2$
and $\xi=0.1$.
}
\end{figure}
In this talk,
a method is reviewed for calculating the 
GPDs of hadrons made of composite constituents,
in an Impulse Approximation (IA) framework. 
In this scheme, GPDs are given by convolutions
between the light cone non-diagonal momentum distribution of the hadron
and the GPD of the constituent. 
Results are presented for the nucleon and for the $^3$He nucleus.
The simplest GPD is the unpolarized one, $H_q(x,\xi,\Delta^2)$.
Usually one works in a system of coordinates where
the photon 4-momentum, $q^\mu=(q_0,\vec q)$, and $\bar P=(P+P')/2$ 
are collinear along $z$,
$\xi$ is the so called ``skewedness'', defined
by the relation 
$
\xi = - {n \cdot \Delta / 2} = - {\Delta^+ / 2 \bar P^+}
= { x_{Bj} /( 2 - x_{Bj}) } + {{O}} \left ( {\Delta^2 / Q^2}
\right ) ~,
$
where $n$
is a light-like 4-vector
with $n \cdot \bar P = 1$.
One should notice that $\xi$
can be fixed experimentally.
The constraints of $H_q(x,\xi,\Delta^2)$ are: 
i) the 
``forward'' limit, 
$\Delta^2=\xi=0$, where one 
recovers the usual parton distribution;
ii)
the integration over $x$, giving the contribution
of the quark of flavor $q$ to the Dirac 
form factor;
iii) the polynomiality property,
involving higher moments of GPDs.
In Ref. \cite{io3},
an IA expression
for $H_q$ of a given hadron target $A$
has been obtained.
Assuming that the interacting parton belongs
to a bound constituent $N$ 
with momentum $p$ and removal energy $E$,
for small values of $\xi^2$ and
$\Delta^2 \ll Q^2,M^2$, it reads:
\begin{eqnarray}
\label{flux}
H_q^A(x,\xi,\Delta^2) 
& =  & 
\sum_N \int dE \int d \vec p
\, 
P_{N}^A(\vec p, \vec p + \vec \Delta, E )
{\xi' \over \xi}
H_{q}^N(x',\xi',\Delta^2)~.
\label{spec}
\end{eqnarray}
In the above equation, the kinetic energies of the 
residual system and of the recoiling target have been neglected, 
$P_{N}^A (\vec p, \vec p + \vec \Delta, E )$ is
the one-body off-diagonal spectral function
for the constituent $N$ in the target $A$,
the quantity
$
H_q^N(x',\xi',\Delta^2)
$
is the GPD of the bound constituent 
up to terms of order $O(\xi^2)$, and 
$
\xi'  =  - \Delta^+ / 2 \bar p^+~,
$
$ x' = (\xi' / \xi) x$~.
In Ref. \cite{io3}, it is discussed that
Eq. (\ref{spec}) fulfills the constraints $i)-iii)$ listed above.
This  formalism has been applied in Ref \cite{io3} 
to the nucleon target. The spectral function of the composite constituent
quarks has been approximated by a momentum distribution, calculated
within the Isgur and Karl model \cite{ik},
convoluted with the GPDS of the constituent quarks themselves.
The latter are modeled by using the structure functions of the constituent
quark, obtained generalizing to the GPDs case the approach of \cite{io6}
which is, in turn, built following Ref \cite{acmp},
the double distribution representation of GPDs \cite{rad1},
and a recently proposed constituent quark 
form factor \cite{sim}.
In ref. \cite{iosavv}, an analysis has been done to
show that, in calculating parton distributions 
from quark models, as it is done here,
the role played by the structure
of the constituent quarks is different from that arising
from relativistic effects. 
\begin{figure}[ht]
\centerline{\epsfxsize=3.0in\epsfbox{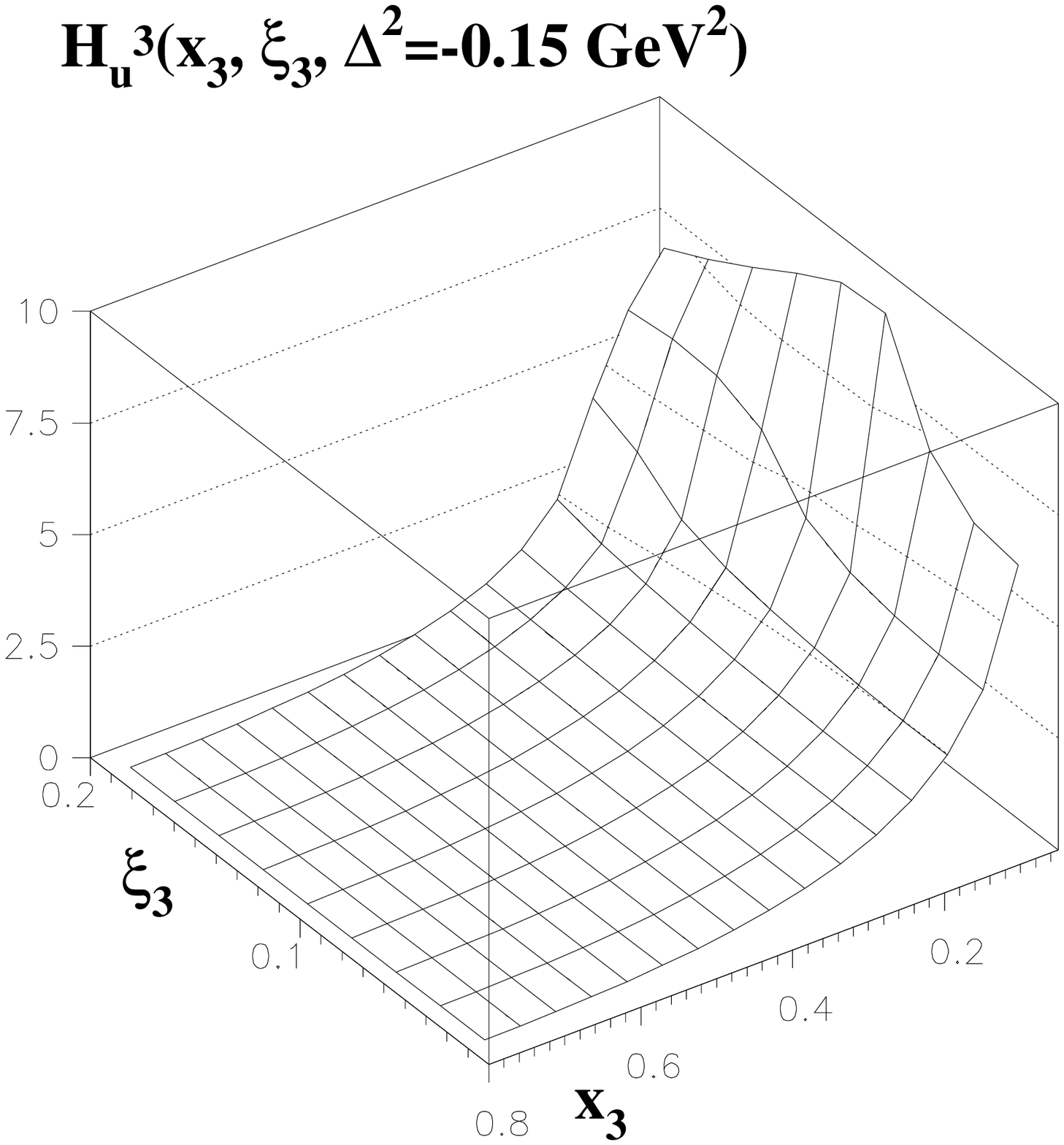}
\epsfxsize=3.0in\epsfbox{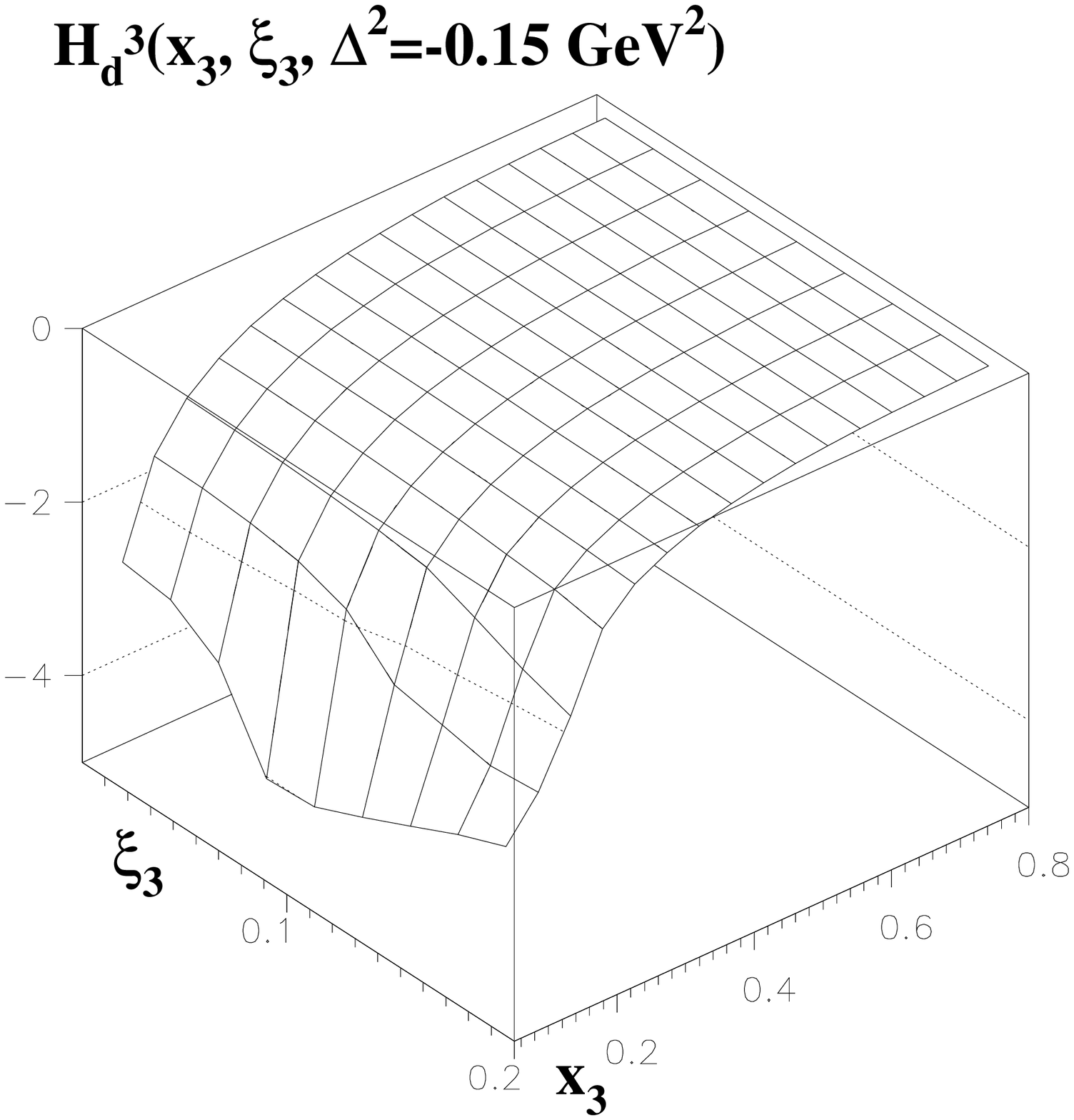}}   
\caption{
The GPD $H_q^3$, for
$\Delta^2 = -0.15$ GeV$^2$, as a function
of $x_3 = 3 x$, for the allowed values of $\xi_3=3 \xi$.
Results for the flavors $u$ and $d$ are shown in 
the left and right panels, respectively.}
\end{figure}
In particular, the structure 
parameterized following Ref. \cite{acmp},
improves the description of the low-x behavior, while
relativistic effects governs 
the high-$x$ tail of the data. A typical example
of this trend is shown in the left panel
of Fig.1. 
Details can be found in Ref. \cite{iosavv}.
Results for GPDs have been discussed in Ref. \cite{io3} for the 
helicity-independent and helicity independent GPDs.
The model has been built to be valid
at the so-called hadronic scale, $\mu_0^2=0.34$ GeV$^2$,
and in the unpolarized case also
the NLO QCD evolution of the results up to experimental scales
has been discussed.
In the right panel of Fig.1, typical results
are shown for $H_q(x,\xi,\Delta^2$).
This approach permits to access the so-called ERBL region,
difficult to study within quark models.
To this aim, another approach has been recently proposed,
adding a meson cloud to a light-front quark model scenario \cite{pb}.

Let us discuss now the GPDs for nuclear few-body systems.
The deuteron target has been studied 
carefully \cite{cano1,cano2}. 
In the coherent, no-break-up channel,
at low values of $\Delta^2$, the differential cross section
for exclusive photon production off the deuteron target has been
found to be comparable
to that obtained for the proton target, demonstrating the possibility
of its measurement.
The study of GPDs for $^3$He is very interesting, since
for $^3$He realistic studies 
are possible, so that conventional effects
can be distinguished from the exotic ones.
Besides, $^3$He is an effective 
polarized free neutron \cite{friar} and
it will be a natural target to study the
helicity-dependent GPDs of the free neutron. 
In what follows,
the results will be reviewed of an IA
calculation of $H_q^3$, Eq. (\ref{spec}) 
for the quark of flavor $q$ of
$^3$He \cite{prc}.
Use has been made of
a realistic non-diagonal spectral function,
so that Fermi motion and binding effects are rigorously estimated.
The scheme is valid for 
$\Delta^2 \ll Q^2,M^2$
and it permits to
calculate GPDs in the kinematical range relevant to
the coherent channel of DVCS off $^3$He.
$H_q^3$, 
has been evaluated in the nuclear Breit Frame.
The non-diagonal spectral function appearing in
Eq. (\ref{spec}) has been calculated 
along the lines of Ref. \cite{gema},
by means of a AV18 wave functions.
The other ingredient in Eq. (\ref{spec}), i.e.
the nucleon GPD $H_q^N$, has been modelled in agreement with
the Double Distribution representation \cite{rad1}.
In this model, whose details are found in Ref. \cite{prc},
the $\Delta^2$-dependence of $H_q^N$ is given by
$F_q(\Delta^2)$, i.e. the contribution
of the quark of flavor $q$ to the nucleon form factor.
Typical numerical results are shown in Fig. 2.
Anyway, the main result of this investigation
is not the size and shape of the obtained $H_q^3$ for $^3$He,
but the nature of nuclear effects on it.
This permits to test 
the accuracy of prescriptions
proposed to estimate nuclear GPDs \cite{cano2}.
In \cite{prc} nuclear effects are thoroughly discussed.
Some general trends of them can be summarized as follows:
i) nuclear effects, for $x_3 \leq 0.7$, are as large as 15 \% at most;
ii) Fermi motion and binding have their main effect
for $x_3 \leq 0.3$, at variance with what happens
in the forward limit;
iii) nuclear effects increase with
increasing $\xi$ and
$\Delta^2$, for $x_3 \leq 0.3$;
iv) nuclear effects for the $d$ flavor are larger than
for the $u$ flavor.
In general, it is found that the realistic calculation
yields a rather different result with respect to
a simple parameterizations
of nuclear GPDs, as some of the ones proposed
in Ref. \cite{cano2}.
In Ref. \cite{gron}, it is shown that nuclear effects
are found to 
depend also on the choice of the NN potential, 
at variance with what happens in the forward case.
The study of nuclear GPDs turns out therefore to be very 
fruitful, being able to detect
relevant details of the nuclear structure at short light-cone distances.
The obtained $^3$He GPDs have now to be used 
to estimate cross-sections. Calculations for the helicity
dependent channels are in progress.

\end{document}